\documentclass[11pt]{article}
\usepackage{amsmath, amssymb, graphics}

\usepackage{graphics, psfrag,color}
\usepackage{graphicx}

\def\id{\protect{{1 \kern-.28em {\rm l}}}}

\def\be{\begin{eqnarray}}
\def\ee{\end{eqnarray}}
\def\p{{\partial}}
\def\nn{\nonumber}
\makeatletter
\renewcommand\section{\@startsection {section}{1}{\z@}%
								   {-3.5ex \@plus -1ex \@minus -.2ex}%
								   {2.3ex \@plus.2ex}%
								   {\normalfont\large\bfseries}}
\renewcommand\subsection{\@startsection{subsection}{2}{\z@}%
								   {-3.25ex\@plus -1ex \@minus -.2ex}%
								   {1.5ex \@plus .2ex}%
								   {\normalfont\normalsize\bfseries}}

\makeatother

\newcommand{\mathsym}[1]{{}}
\def\k{\kappa}

\def\tk{{\text{k}}}
\def\p{{\partial}}
\def\nn{\nonumber}

\def\dalemb#1#2{{\vbox{\hrule height .#2pt
		\hbox{\vrule width.#2pt height#1pt \kern#1pt
				\vrule width.#2pt}
		\hrule height.#2pt}}}

\def\half{{\textstyle{1\over2}}}
\let\a=\alpha \let\b=\beta \let\g=\gamma \let\d=\delta \let\e=\epsilon
\let\z=\zeta \let\h=\eta \let\th=\theta  \let\k=\kappa
\let\l=\lambda \let\m=\mu \let\n=\nu \let\x=\xi \let\p=\pi 
\let\s=\sigma \let\t=\tau   \let\c=\chi 
 \let\vep=\varepsilon
\let\w=\omega	    \let\D=\Delta \let\Th=\Theta \let\L=\Lambda
 \let\P=\Pi \let\S=\Sigma  
\let\C=\Chi 
\let\la=\label \let\ci=\cite 
  
\def\nn{\nonumber} \def\bd{\begin{document}} \def\ed{\end{document}}
\def\ds{\documentstyle} \let\fr=\frac \let\bl=\bigl \let\br=\bigr
\let\Br=\Bigr \let\Bl=\Bigl
\let\bm=\bibitem
\let\na=\nabla
\def\tU{{\widetilde U}}
\let\pa=\partial \let\ov=\overline
\def\ie{{\it i.e.\ }}
\def\ba{\begin{array}}
\def\ea{\end{array}}
\def\ft#1#2{{\textstyle{{\scriptstyle #1}\over {\scriptstyle #2}}}}
\def\fft#1#2{{#1 \over #2}}
\def\F#1#2{{ F_{#1}^{(#2)} }}
\def\cF#1#2{{ {\cal F}_{#1}^{(#2)} }}

\def\R{{\bf R}}
\def\sst#1{{\scriptscriptstyle #1}}
\def\oneone{\rlap 1\mkern4mu{\rm l}}
\def\e7{E_{7(+7)}}
\def\td{\tilde}
\def\wtd{\widetilde}
\def\im{{\rm i}}
\newcommand{\ho}[1]{$\, ^{#1}$}
\newcommand{\hoch}[1]{$\, ^{#1}$}
\newcommand{\bea}{\begin{eqnarray}}
\newcommand{\eea}{\end{eqnarray}}
\newcommand{\ra}{\rightarrow}
\newcommand{\lra}{\longrightarrow}
\newcommand{\Lra}{\Leftrightarrow}
\newcommand{\ap}{\alpha^\prime}
\newcommand{\bp}{\tilde \beta^\prime}
\newcommand{\cB}{{\cal B}}
\newcommand{\cO}{{\cal O}}
\newcommand{\vecx}{\vec{x}}
\newcommand{\vecy}{\vec{y}}
\newcommand{\vecp}{\vec{p}}
\newcommand{\vecq}{\vec{q}}
\newcommand{\tr}{{\rm tr} }
\newcommand{\Tr}{{\rm Tr} }

\newcommand{\cL}{{\cal L}}
\newcommand{\cA}{{\cal A}}
\newcommand{\cD}{{\cal D}}
\def\sst#1{{\scriptscriptstyle #1}}

\def\ve{\varepsilon}
\def\vf{\varphi}
\def\F{\Phi}
\def\wg{\wedge}

\newcommand{\wt}{\widetilde}
\newcommand{\oh}[1]{{\cal O}( #1 )}
\newcommand{\largeoh}[1]{{\cal O}\left( #1 \right)}
\def \foot {\footnote}
\def \bi{\bibitem}

\def \tr {{\rm tr}}
\def \ha {{1 \over 2}}
\def \td {\tilde}
\def \ci{\cite}
\def \N {{\mathcal N}}
\def \const {{\rm const}}
\def \ss {\sum\limits_{i=1}^3 }
\def \t {\tau}
\def\S{{\mathcal S} }
\def \nn {\nu}
\def \XX {{\rm X}}

\def \lra {\leftrightarrow}
\def \vom {{\bar \omega}}
\def \E {{\mathcal	E}} \def \J {{\mathcal	J}}
\def \YY {{\rm Y}}

\def \d {\del}
\def \rJ {{J}}
\def \sms {sigma models\ }
\def \sm {sigma model\ }
\def \L {\Lambda}
\def \gl {\ell}
\def \tr {{\rm tr\ }}
\def\z{\zeta}
\def\zi{\zeta_1}
\def\zii{\zeta_2}
\def\K{\mbox{K}}
\def\eE{\mbox{E}}	\def \vt {\vartheta}
\def \vr {\varrho}
\def \wup {w}

\def\dg{\dagger}
\def\a{\alpha}
\def\b{\beta}
\def\e{\varepsilon}
\def\p{\phi}
\def\ap{\alpha^\prime}
\def\I{{\cal I}}

\def\R{{\bf R}}
\def\Z{{\bf Z}}
\def\C{{\bf C}}
\def\P{{\bf P}}
\def\xb{{\bar X}}
\def\Tr{{\rm  Tr}}
\def\tr{{\rm  tr}}

\def \bs {\bar \s}
\def \btau {\bar \tau}

\def \del{\partial}
\def \a {\alpha}
\def \aa {{\a'}}
\def\g{\gamma}
\def\s{\sigma}
\def\z{\zeta}
\def\zi{\zeta_1}
\def\zii{\zeta_2}
\def\ov{\over}

\def\I{{\cal I}}
\def\J{{\mathcal J}}
\def \ok {{1\ov \k}}
\def\LL{{\mathcal L }}
\def \jL {{J}}
\def \om {\omega}
\def \cL {{\mathcal L}} \def \cH {{\mathcal H}}
\def\E{{\mathcal E}}
\def\b{\beta}
\def\l{\lambda}
\def\eps{\epsilon}
\def\vep{\varepsilon}
\def \De {{\mathcal D}}

 \def \cV {{\cal V}}
\def  \Jt {	 {J}_{\rm tot}	  }

\def \k {\kappa}
\def\foot{\footnote}
\def \four{{\textstyle {1\ov 4}}}
 \def \third { \textstyle {1\ov 3
}}
\def\det{\hbox{det}}
\def \ci {\cite}

\def \foot {\footnote}
\def \bi{\bibitem}

\def \tr {{\rm tr}}
\def \ha {{1 \over 2}}
\def \tid {\tilde}
\def \vv {{\rm v}}
\def \tl {{\tilde \l}}
\def \XX {{\rm X}}
\def \ta {{\tilde \a}}
\def \fo { {1\ov 4}}
\def \ep {\epsilon}
\def \inti {{\int^{2\pi}_0 {d \sigma \ov 2 \pi}}}

\def \d {\partial}
\def \K {{\rm S}}
\def \el {\ell}
\def \Tr {{\rm Tr}}
\def \P {\Phi}
\def \l	 {\lambda}
\def \tl {{\tilde \l}}
\def \bl {{\tilde \l}}
\def \const {{\rm const}}
\def \V {v}

\def \bv {v^*}
\def \vv {{\rm v}}
\def \LL {{\mathcal L}}
\newcommand{\PV}[1]{P_{\!\!_{V_{#1}}}}

\def \bL {\ell}
\def \M {{\mathcal M}}
\def \N {{\mathcal N}}
\def \S {{\rm S}}
\def \vn {\vec n}
\def \tl {\td \l}
\def \td {\tilde}
\def \Prod {\Pi}
\def \O {{\mathcal O}}
\def \Q {{\rm  Q}}
\def \D {\Delta}
\def \N {{\mathcal N}}
\def\tN{{\tilde N}}

\def \m {\mu}
\def \vs {\vec \s}
\def \ie {i.e.}

\def \cD {{\cal D}}

\def  \le  {\l_{\rm eff}}

\def\as{{\a}}
\newcommand{\bra}[1]{\mbox{$\langle #1 |$}}
\newcommand{\ket}[1]{\mbox{$| #1 \rangle$}}

\def\thb{\bar{\theta}}
\def\Thb{\bar{\Theta}}
\def\barp{\bar{p}}
\def\barq{\bar{q}}
\def\barc{\bar{c}}
\def\bard{\bar{d}}
\def\e{\epsilon}

\def \bi{\bibitem}
\def \la {\label}

\def \l {\lambda}
\def\foot{\footnote}
\def \tl  {{\tilde \l}}
\def \sql {{\sqrt \l}}
\def \adss {$AdS_5 \times S^5$\ }
\newcommand{\rf}[1]{(\ref{#1})}
\def \ov {\over}

\def\th{\theta}
\def\Th{\Theta}
\def\vth{\vartheta}

\def\vth{\vartheta}
\def\ra{\rightarrow}
\def\N{{\cal N}}
\def\F{{\cal F}}
\def\cc{\circ}
\def\eqv{\equiv}

\def\ni{\noindent}

\def \cT {{\cal T}}
\def \no {\nonumber}
\def \del {\partial}
\def \E {{\cal E}}
\def \S {{\cal S}}
\def \J {{\cal J}}
\def \bi{\bibitem}
\def \la {\label}
\def\foot{\footnote}

\def \adss {$AdS_5 \times S^5$\ }
\def \arccot {{\rm arccot}}

\def\pic #1#2{\hbox{\lower#1pt\hbox{~\mbox{\epsfxsize=20truemm \epsffile{#2}}}}}

\def\pic #1#2#3{\hbox{\lower#1pt\hbox{~\mbox{\includegraphics[scale=#3]{#2}}}}}

\def \bt {\bar\theta}
\def \te {\theta}

\def \cc {{\rm f}}
\def \d {\delta}

\def \cL {{\cal L}}
\def \S	 {{\cal S}}
\def \pp {{q}}
\def \vt {\vartheta}
\def \mm {{\cal	 \ell}}
\def \Z {{\cal Z}}
\def \pa {\partial}
\def \C {{\cal C}}
\def \be {\bea}
\def \ee {\eea}
\def \c {\gamma}  \def \d {\delta}
\def \eps {\epsilon}

\def \bp {\begin{pmatrix}}	\def \ep {\end{pmatrix}}

 \def \T {{\cal T}}

\def \bp {\begin{pmatrix}}	\def \epm {\end{pmatrix}}
\def \ha {{\textstyle{1 \ov 2}}}
\def \r {\rho}
\def \S {{\cal S}}
\def \z {\chi}

\def \sn {{\rm sn}}
\def \dn {{\rm dn}}
\def \cn {{\rm cn}}
\def \am {{\rm am}}

\def \const { {\rm const} }
\def \k {\kappa}
\def \t {\tau}
\def \acosh {{\rm arccosh \,}}
\def \AdSS {${\rm AdS}_5 \times {\rm S}^5 \ $}
\def \adss {${\rm AdS}_5 \times {\rm S}^5 $}
\def \AdS {${\rm AdS}_5 \ $}
\def \ads {${\rm AdS}_5 $}
\def \size {\displaystyle}
\def \bei {\begin{itemize}}
\def \eei {\end{itemize}}
\def \RS {$R_t \times S^5$}
\def \half {\frac{1}{2}}
\def \pih {\frac{\pi}{2}}
\def \kpih {\frac{\k\pi}{2}}
\def \sql {\sqrt{\l}}

\def \z {\xi}
\def \x {\chi}

\def \bM {\mathbb{M}}
\def \bC {\mathbb{C}}
\def \bS {\mathbb{S}}
\def \bK {\mathbb{K}}
\def \bE {\mathbb{E}}

\def \beq{\be}
\def \eeq{\ee}
\def \I {{\rm I}}
\def \II {{\rm II}}
\def \de {\del}

\def\Ra {\Rightarrow}
\def\tk {\td \k}
\def\tell {\td\ell}
\def \piho {\frac{\pi_0}{2}}
\def \ktk {\frac{\k}{\tk}}

\def \P {{\cal P}}

\begin{document}


\overfullrule=0pt
\parskip=2pt
\parindent=12pt
\headheight=0in \headsep=0in \topmargin=0in \oddsidemargin=0in

\vspace{ -3cm}
\thispagestyle{empty}
\vspace{-1cm}

\rightline{Imperial-TP-EM-2011-01}

\begin{center}
\vspace{1cm}
{\Large\bf

Splitting of folded strings in AdS$_3$
}

\vspace{.2cm}

 E. M. Murchikova\footnote{e.murchikova@imperial.ac.uk}

\vskip 0.6cm

{\em
Blackett Laboratory, Imperial College,
London SW7 2AZ, U.K.
\\
\vskip 0.08cm
Skobeltsyn Institute, Moscow State University, Moscow, 119991, Russia
 }

\vspace{.2cm}

\end{center}

\begin{abstract}

In this paper we present semiclassical computations of the splitting
of folded spinning strings in AdS$_3,$ which may be of interest in
the context of AdS/CFT duality. We start with a classical closed
string and assume that it can split into two closed string
fragments, if at a given time two points on it coincide in target
space and their velocities agree. First we consider the case of the
folded string with large spin. Assuming the formal large-spin
approximation of the folded string solution in AdS$_3,$ we can
completely describe the process of splitting: compute the full set
of charges and obtain the string solutions describing the evolution
of the final states. We find that, in this limit, the world surface
does not change in the process and the final states are described by
the solutions of the same type as the initial string, i.e. the
formal large-spin approximation of the folded string in AdS$_3.$
Then we consider the general case --- splitting of string given by
the exact folded string solution. We find the expressions for the
charges of the final fragments, the coordinate transformations
diagonalizing them and, finally, their energies and spins. Due to
the complexity of the initial string profile, we cannot find the
solutions describing the evolution of the final fragments, but we
can predict their qualitative behavior. We also generalize the
results to include circular rotations and windings in S$^5.$

\end{abstract}

\setcounter{equation}{0}
\setcounter{footnote}{0}
\setcounter{section}{0}

\newpage

\section{Introduction}

\renewcommand{\theequation}{1.\arabic{equation}}
 \setcounter{equation}{0}

Decay properties of massive strings have been studied for a long
time \cite{Green,Mitchell,Dai,Okada,Sundborg,wilkinson,Turok,
Amati2,IK,Manes,Iengo:2002tf,IR,IR2,Plefka,Janik}. In this paper we
present semiclassical computations of the splitting of folded
spinning strings in AdS$_3.$
 Classical string solutions have
proved to be a useful tool for exploring the AdS/CFT correspondence
in the sector of large charges \cite{bmn,gkp,ft1,ft2,tov,ple,19,ar}.

For flat Minkowski space splitting of semiclassical strings was
analyzed in detail in \cite{IR,IR2}, for \RS \ space in
\cite{Plefka,Janik}. There is an obvious lack of results in AdS
space, and the purpose of the present paper is to fill this gap.
 Following the conventional approach,
we start with a classical closed string and assume that it can split
into two fragments, if at a given time $\t_0$ two points on it
coincide in target space and their velocities agree. Closed string
periodicity conditions are separately imposed on each of the two
final pieces. Initial conditions are defined by the initial string
at $\t_0.$ The relations between the energies and spins of the cut
fragments --- together with ``conservation laws'' of splitting
$E(E_\I,E_\II,...),$ $S(S_\I,S_\II,...),$ etc --- are completely
determined by the charge conservation. Thus they may be found (at
least parametrically) for the initial string solution of arbitrary
complexity. Determining the evolution is much more complicated:
one has to solve the string equations
with the boundary conditions given by a part of the profile of the
initial string. At the moment, this is possible only in the simplest cases.

The main purpose of this paper is to investigate
splitting of folded spinning string in AdS$_3$ \ci{gkp}
\be\la{i1}
\ba{c}
\size
	Y_0+i Y_5 = \dn\big[{\kappa \ell^{-1}  \sigma},\ -\ell^2\big] \ e^{i\k \t},
	\qquad
	Y_1+i Y_2 = \ell \ \sn\big[{\kappa \ell^{-1} \sigma},\ -\ell^2\big]	 \ e^{i\w \t},
\\
\size
	\kappa=\frac{2}{\pi} \ell \ \bK[-\ell^2],
	\qquad
	\frac{w^2}{\kappa^2}= 1+ \frac{1}{\ell^2},
\ea
\ee
where $\sn[z,m]$ and $\dn[z,m]$ are Jacobi elliptic functions,
$\bK[z]$ is the complete elliptic integral of
the first kind. First we consider the limit of the folded string with large spin.
Then solution \rf{i1} may be approximated by
\be\la{i2}
\ba{l}
	 \size Y_0+i Y_5= \cosh (\k \s) \, e^{i \w \t},
	 \qquad
	 \size Y_1+i Y_2= \sinh (\k \s) \, e^{i \w \t},
	 \qquad
	 \k=\w \gg 1.
\ea \ee In this simple case, we can completely describe the process
of splitting: compute the full set of charges and find string
solutions describing the evolution of the final states. It appeared
that when such a string splits, the world surface does not change in
the process and the final states are described by the solutions of
the same type as \rf{i2}: \be\la{i3}
	Y_{\I,\II \, 0} + i Y_{\I,\II \, 5} = \cosh (\k_{\I,\II} \s) e^{i \k_{\I,\II} \t}, \quad
	Y_{\I,\II \, 1} + i Y_{\I,\II \, 2} = \sinh (\k_{\I,\II} \s) e^{i \k_{\I,\II} \t}, \quad
	\k_{\I,\II}=\k \frac{\pi \mp 2\s_0}{2\pi},
\ee
where $\s_0$ parameterizes the coordinate of the splitting point.

In the general case we find expressions for
the charges of the final fragments, the
coordinate transformations that diagonalize them
and, at the end, their energies and spins
as the functions of $\ell$ and $\s_0$ (in
the coordinate system where no non-Cartan components
present). These are
\be\la{i4}
\ba{c}
\size
	E_{\I,\II} =\frac{\sql}{2} \sqrt{ (\k \bC_{\I,\II} + \w \bS_{\I,\II})^2 - \bM_{\I,\II}^2 (\w + \k)^2}
		+ \frac{\sql}{2} \sqrt{ (\k \bC_{\I,\II} - \w \bS_{\I,\II})^2 - \bM_{\I,\II}^2 (\w - \k)^2}
	\\
\size
	S_{\I,\II} = \frac{\sql}{2} \sqrt{ (\k \bC_{\I,\II} + \w \bS_{\I,\II})^2 - \bM_{\I,\II}^2 (\w + \k)^2}
		- \frac{\sql}{2} \sqrt{ (\k \bC_{\I,\II} - \w \bS_{\I,\II})^2 - \bM_{\I,\II}^2 (\w - \k)^2}.
\ea
\ee
Here
\be\la{i5}
\ba{c}
\size
	\k \bC_{\I,\II} =
	\half \E_{\rm fold}
	\mp \frac{\ell}{\pi} \bE\left[ \am[\k \ell^{-1} \s_0,-\ell^2],-\ell^2\right],
\\[8pt]
\size
	\w\bS_{\I,\II} = \half \S_{\rm fold}
	\mp \sqrt{1+\ell^2} \left( -\frac{2}{\pi} \s_0 \bK[-\ell^2] +
	\bE\left[ \am[\k \ell^{-1} \s_0,-\ell^2],-\ell^2\right] \right),
\\[8pt]
\size
	\bM_{\I,\II} = \pm \frac{\ell^2}{\k \pi} \ \cn\left[\k \ell^{-1} \s_0,-\ell^2\right],
\ea
\ee
where $E_{\rm fold}=\sql \E_{fold}$ and $S_{fold}=\sql \S_{\rm fold}$
are the energy and spin of the folded string \rf{i1};
$\bE[z]$ and $\bE[z,m]$ are the complete and incomplete elliptic integrals of
the second kind, respectively, and $\cn[z,m]$ is a Jacobi
elliptic function.
These relations parametrically encode the conservation laws of
splitting, namely $E(E_\I,E_\II),$ $S(S_\I,S_\II),$ etc.

Due to the complexity of the folded string profile \rf{i1}, we are
unable to find the solutions describing the evolution of the
final fragments explicitly. However, we can describe the evolution
qualitatively. Let us examine the case of large but not infinitely
large (as in \rf{i2}) spin, with the cut occurring far enough from the
string ends for $\s_0$ to satisfy $\k(\pi/2 - \s_0)\gg 1.$ In this
limit one expects the final pieces to have almost the standard folded
shape \rf{i1}, disturbed by a kink moving along the string, similar
to the one observed in flat Minkowski space \ci{IR}. The kink is a
``correction'' to the ``leading'' folded shape of the cut fragments,
thus the angle of bending has to depend on the position of the kink.
It may be substantial at the string ends but must be
small close to the center.

The results obtained for the folded string in AdS$_3$
generalizes to include
circular rotations and windings in S$^5.$
We discuss such a generalization with the example of the string in
AdS$_3\times$S$^3.$

\vspace{20pt}

The rest of the paper is organized as follows. In Section 2 we
introduce notations and discuss a general approach to studying
splitting of classical bosonic closed strings in AdS$_5\times$S$^5.$
Section 3 is a review of the splitting of the folded strings in flat
Minkowski space. Section 4 is dedicated to the splitting of
Gubser--Klebanov--Polyakov folded strings in AdS$_3.$ The results
obtained in AdS$_3$ are generalized to include circular rotations
and windings in S$^5$ in Section 5.

\section{Splitting of closed strings in AdS$_5 \times$S$^5$. General formalism.}

\renewcommand{\theequation}{2.\arabic{equation}}
 \setcounter{equation}{0}

In this section we discuss a general approach to
studying of splitting of classical closed bosonic strings
in AdS$_5 \times$S$^5$.

The action for a bosonic string
in AdS$_5 \times$S$^5$ reads
\begin{equation}\label{1}
	I_B=\frac{1}{2}
		T \int d\tau \int \limits^{2\pi}_0 d\sigma (L_{AdS}+L_S),
	\qquad T=\frac{R^2}{2\pi \alpha'}=\frac{\sqrt{\lambda}}{2\pi},
\end{equation}
where
\begin{equation}\label{1.1}
	L_{AdS}=-\partial_a Y_P \partial^a Y^P-\tilde{\Lambda} (Y_P Y^P+1) ,
	\qquad
	L_{S}=-\partial_a X_M \partial^a X_M+\Lambda (X_M X_M-1).
\end{equation}
Here $X_M, \, M=1,...,6$ and $Y_P, \, P=0,...,5$ are embedding coordinates
of $R^6$ with the Euclidean metric $\delta_{MN}=(+1,+1,+1,+1,+1,+1)$
in $L_S$ and of $R^{2,4}$ with
$\eta_{PQ}=(-1,+1,+1,+1,+1,-1)$ in $L_{AdS},$ respectively ($Y_P=\eta_{PQ}Y^Q$).
$\Lambda$ and $\tilde{\Lambda}$ are the Lagrange multipliers imposing the two
hypersurface conditions:
\be\la{hypcon}
	\eta_{PQ} Y^P Y^Q = -1 \qquad X_M X_M = 1.
\ee
The action (\ref{1}) is supplemented with the conformal gauge constraints
\begin{equation}\label{3}
	\dot{Y}_P \dot{Y}^P+Y'_P Y'^P+\dot{X}_M \dot{X}_M+ X'_M X'_M =0, \qquad
	\dot{Y}_P Y'^P+\dot{X}_M X'_M=0
\end{equation}
and the closed string periodicity conditions
\begin{equation}\label{6}
	Y_P(\tau,\sigma+2\pi)=Y_P(\tau,\sigma), \qquad X_M(\tau,\sigma+2\pi)=X_M(\tau,\sigma).
\end{equation}
The classical equations of motion following from (\ref{1}) are
\begin{equation}\label{2}
	\begin{array}{lll}
		\partial^a \partial_a Y_P-\tilde{\Lambda} Y_P=0,	& \tilde{\Lambda}=\partial^a Y_P \partial_a Y^P,  & Y_P Y^P=-1, \\
		\partial^a \partial_a X_M+{\Lambda} X_M=0,	& {\Lambda}=\partial^a X_M \partial_a X_M,	& X_M X_M=1.
	\end{array}
\end{equation}
The action is invariant under the $SO(2,4)$ and $SO(6)$ rotations with correspondent conserved
(on-shell) charges
\be\la{char}
	S_{PQ}= \sqrt{\lambda} \int\limits^{2\pi}_0 \frac{d\s}{2\pi}
	(Y_P \dot{Y}_Q-Y_Q \dot{Y}_P),
	\qquad
	J_{MN}= \sqrt{\lambda} \int\limits^{2\pi}_0 \frac{d\s}{2\pi}
	(X_M \dot{X}_N-X_N \dot{X}_M)
	\ .
\ee
We will be working with ``spinning'' string solutions
which have nonzero values of these charges.

It is useful to solve the constraints \rf{hypcon}
by choosing an explicit parametrization of the embedding
coordinates $Y_P$ and $X_M,$ e.g.
\begin{equation}	\label{an1}
\ba{c}
\size
	Y_{05}= Y_0+iY_5 =\cosh{\r} e^{i t} \ ,
	\\[8pt]
\size
	Y_{12}= Y_1+iY_2 =\sinh{\r} \cos \te e^{i \phi_1}\ ,  \qquad
	Y_{34}= Y_3+iY_4 =\sinh{\r} \sin \te e^{i \phi_2} \ ;
\ea
\end{equation}
\begin{equation}	\label{X1}
\ba{c}
\size
	X_{12}=X_1+iX_2 = \sin{\g} \cos \psi e^{i \varphi_1} \ , \qquad
	X_{34}=X_3+iX_4 = \sin{\g} \sin \psi e^{i \varphi_2} \ ,
	\\[8pt]
\size
	X_{56}=X_5+iX_6 = \cos{\g} e^{i \varphi_3} \ .
\ea
\end{equation}
The corresponding metrics take the form
 \begin{equation}\label{io}
 ds^2_{AdS_5} = - \cosh^2 \rho \ dt^2 + d\rho^2	 + \sinh^2 \rho \ (d \theta^2 + \cos^2 \theta \ d \phi_1^2 +
 \sin^2 \theta \ d \phi_2^2)  \
 \end{equation}
  \begin{equation}\label{io2}
 ds^2_{S^5} = \cos^2 \g \ d\varphi_3^2 + d\g^2 + \sin^2 \g \ (d \psi^2 + \cos^2 \psi \ d \varphi_1^2 +
 \sin^2 \psi \ d \varphi_2^2).
 \end{equation}
The Cartan generators of $SO(2,4)$ corresponding to the three linear
isometries of the \AdS metric are the translations in the AdS-time
$t$ and two angles $\phi_1$ and $\phi_2:$
\be \la{EESS}
S_0 \equiv
S_{05} \equiv E = \sqrt{\lambda} \E, \quad S_1 \equiv S_{12} =
\sqrt{\lambda} \S_1, \quad S_2 \equiv S_{34} = \sqrt{\lambda} \S_2 .
\ee
The Cartan generators of $SO(6)$ corresponding to the three
linear isometries of the $S^5$ metric are the translations in the
three angles $\varphi_1,$ $\varphi_2$ and $\varphi_3:$
\be \la{JJJ}
J_1 \equiv J_{12} = \sqrt{\lambda} \J_1, \quad J_2 \equiv
J_{34} =
\sqrt{\lambda} \J_2, \quad J_3 \equiv J_{56} = \sqrt{\lambda} \J_3.
\ee


Let us consider a string solution
\be\la{gin}
	X_M = {X_{in}}_M(\t,\s), \qquad Y_P={Y_{in}}_P(\t,\s)
\ee with energy $E$ and spins $S_n, \ J_k. $ We assume that if at a
given $\t_0$ two points on the string coincide in the target space
\be
	 {X_{in}}_M(\t_0,\s_1) = {X_{in}}_M(\t_0,\s_2) \qquad
	 {Y_{in}}_P(\t_0,\s_1) = {Y_{in}}_P(\t_0,\s_2)
\ee
and their velocities agree
\be
	 {\dot X_{in \, M}}(\t_0,\s_1) = {\dot X_{in \, M}}(\t_0,\s_2) \qquad
	 {\dot Y_{in \, P}}(\t_0,\s_1) = {\dot Y_{in \, P}}(\t_0,\s_2),
\ee then the string can split into two pieces \be \la{interval}
	\size {\rm fragment} \ \I \ &:& \s \in (0,\s_1) \cup (\s_2 , 2\pi)\\
\no
	\size {\rm fragment} \ \II &:& \s \in (\s_1, \s_2).
\ee The behavior of the cut fragments is governed by equations
\rf{3} and \rf{2} with the boundary conditions defined by the
initial string at the moment of splitting: \be\la{gin2} \ba{ll}
\size \ba{ll} \size
{X_{\I}}_M (\t_0, \s)={X_{in}}_M(\t_0,\s) \\
\size
{\dot X}_{\I \, M} (\t_0, \s)={\dot X}_{in \, M} (\t_0, \s)
\ea
\qquad \
\ba{ll}
\size
{Y_{\I}}_P (\t_0, \s)={Y_{in}}_P(\t_0,\s) \\
\size
 \dot Y_{\I \, P} (\t_0, \s)= \dot Y_{in \, P} (\t_0, \s)
\ea
\qquad \s \in (0,\s_1) \cup (\s_2 , 2\pi);
\\[15pt]
\size
\ba{ll}
\size
{X_{\II}}_M (\t_0, \s)={X_{in}}_M(\t_0,\s) \\
\size
{\dot X}_{\II \, M} (\t_0, \s)={\dot X}_{in \, M} (\t_0, \s)
\ea
\qquad
\ba{ll}
\size
{Y_{\II}}_P (\t_0, \s)={Y_{in}}_P(\t_0,\s) \\
\size
 \dot Y_{\II \, P} (\t_0, \s)= \dot Y_{in \, P} (\t_0, \s)
\ea
\qquad \s \in (\s_1, \s_2).
\ea
\ee
The closed string periodicity conditions are imposed
on each fragment separately:
\be\la{period}
\ba{c}
\size
	 {X_{\I,\II}}_M(\t,\s)={X_{\I,\II}}_M(\t,\s+2\pi_{\I,\II})\\
\size
	 {Y_{\I,\II}}_P(\t,\s)={Y_{\I,\II}}_P(\t,\s+2\pi_{\I,\II}),\\
\ea
\qquad {\rm where} \qquad
\ba{ll}
\size
	 2\pi_\I=2 \pi - (\s_2 - \s_1) \\
\size
	 2\pi_\II=\s_2 - \s_1.
\ea
\ee

Conditions \rf{gin2} and \rf{period} uniquely determine the final
states. The relations between the energies ($E_{\I,\II})$ and spins
$(S_{\I,\II \, n}, J_{\I,\II \, k})$ of the cut
fragments --- together with ``conservation laws'' of splitting
$E(E_\I,E_\II,...),$ $S(S_\I,S_\II,...),$ etc --- are completely
determined by the charge conservation. Thus they may be found (at
least parametrically) for the initial string solution of arbitrary
complexity. Determining the evolution is much more complicated:
one has to solve the string equations \rf{3}, \rf{2}
with the boundary conditions \rf{gin2} and \rf{period}.
At the moment, this is possible only in the simplest cases.

\section{Splitting of folded strings in the flat space. A review.}\la{sec3}

\renewcommand{\theequation}{3.\arabic{equation}}
 \setcounter{equation}{0}

\def \beq{\be}
\def \eeq{\ee}
\def \I {{\rm I}}
\def \II {{\rm II}}
\def \de {\del}

In this section we review splitting of the folded strings in flat
Minkowski space \ci{IR}. The solution for the folded strings in flat
Minkowski space reads \be\la{In}
	X_0=\ell\tau \ ,
	\quad
	X_1=\r \cos \phi =\ell \cos( \s ) \cos(\tau) \ ,
	\quad
	X_2=\r \sin \phi=\ell\cos(\s )\sin(\tau ).
\ee
The energy and spin
\beq \E= \ell\ , \qquad \J= \half \ell^2 \
\eeq obey the standard Regge relation $\E^2=2 \J.$

\begin{figure}[ht]
\centerline{\includegraphics[scale=0.8]{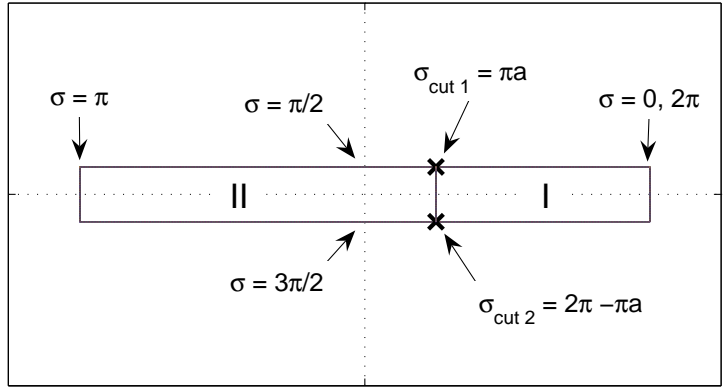}}
\caption{Splitting of the folded string in the flat space.}
\label{fig}\nonumber
\end{figure}

Any two points on the string parameterized by $\s_1$ and $\s_2=2 \pi
- \s_1$ coincide in the target space and their velocities agree at
any given time. Let us assume that at $\tau_0=0 $ the string splits
into two pieces. The cut occurs at $X_1=\ell \cos(a \pi), \ X_2=0,$
i.e. $\s_{cut \ 1}=a \pi$ and $\s_{cut \ 2}=2\pi - a \pi:$
\be\la{frag} \ba{ll}
	\size {\rm fragment} \ \I \	 : \quad \s \in (0,a\pi) \cup (2 \pi - a\pi , 2\pi)\\
	\size {\rm fragment} \ \II: \quad \s \in (a\pi, 2\pi - a\pi),
\ea
\qquad 0 < a < \half.
\ee
Here without loss of generality $0<a<\half,$
i.e. the fragment I is always
``smaller'' than the fragment II (see schematic plot in figure \ref{fig}).

Quantum numbers of the fragment I are
the energy ($\E_\I$), linear momentum ($P_{\I \, i}=\sql \P_{\I \, i}$) and
angular momentum ($\J_\I$):
\be
\E_\I = \P_{\I \, 0} &=&	2
\int\limits_0^{{\pi a} } \frac{d\s}{2\pi}\ \dot X_{0 \, \I} = \ell a\ ,
\\
\P_{\I \, 1} &=& 0\ ,\
\la{px}
\\
\P_{\I \, 2} &=& 2
\int\limits_0^{{\pi a}}\frac{d\s}{2\pi}\ \dot X_{\I \, 2}= 2
\int\limits_0^{{\pi a}} \frac{d\s}{2\pi}\ \cos(\s )= \frac{\ell\sin(\pi a)}{\pi }\ ,
\la{py}
\\
\J_\I={\cal L}_\I + \S_\I&=& 2
\int\limits_0^{\pi a}\frac{d\s}{2\pi}\ (X_{\I \, 1}\dot X_{\I \, 2}-\dot X_{\I \, 1}
X_{\I \, 2})= \ell^2 a \left (\frac{\sin(2\pi a)}{\pi a}-1\right)\ .
\ee
Here the orbital momentum ($L_\I=\sql {\cal L}_\I$)
and spin ($S_\I = \sql {\cal S}_\I$) are\foot{
Orbital momentum is defined as
${\cal L}_\I=X_{cm\I} \ \P_{\I \, 2},$
 where $X_{cm\I}$ is the coordinate of the center of mass of the string
\be\no
X_{cm\I}=\frac{1}{\pi a}
\int\limits_0^{{\pi a}} d\s\ X_{\I \, 1} =\frac{\ell\sin(\pi a)}{\pi a}.
\ee
}
\beq
{\cal L}_\I= \ell^2 a \frac{\sin^2(\pi a)}{(\pi a)^2} \ ,\ \ \ \
\S_\I = \ell^2 a \left(\frac{\sin(2\pi a)}{\pi a}
- \frac{\sin^2(\pi a)}{(\pi a)^2} - 1\right) \ .
\label{orbe}
\eeq
The mass of the fragment I, i.e. its energy in the center-of-mass system
read
\be
M^2_\I= \E_\I^2- \P_\I^2= \ell^2
\left( a^2 - \frac{\sin^2(\pi a)}{\pi ^2} \right) \ .
\ee

The conserved charges for the fragment II may be found similarly:
\be
\ba{c}
\size \E_{\II } = \ell(1-a),\qquad \P_{\II }=-
\frac{\ell\sin(\pi a)}{\pi},\qquad
\J_{\II }={\cal L}_{\II }+\S_{\II },\\
\size
M^2_{\II }= \E_\II^2- \P_\II^2 = \ell^2
\left( (1-a)^2 - \frac{\sin^2(\pi a)}{\pi ^2} \right),\\
\size
{\cal L}_{\II }= \ell^2(1-a) \frac{\sin^2(\pi a)}{(\pi (1-a))^2},
\qquad
\S_{\II}=- \ell^2(1-a)\left (1+
\frac{\sin^2(\pi a)}{(\pi (1-a))^2} + \frac{\sin(2\pi a)}{\pi
(1-a)}\right).
\ea
\ee
The energy, linear momentum and angular momentum
are conserved in the process of splitting:
\beq
\E_\I+\E_\II =\E\ ,\qquad \P_{\I \, 2}+\P_{\II \, 2}=0\ ,\qquad
\J_\I+\J_\II=\J\ .
\eeq

The string solution describing the evolution of the final states may
be found using the general solution for a closed bosonic string in
flat Minkowski space. Imposing the boundary and periodicity
conditions \ci{IR} on it, one finds \be \ba{c} \size
X_{\I \, 0}=2\ell\t,\\
\size
X_{\I \, 1}=\frac{\ell\sin(\pi a)}{\pi a} \left( 1+2\sum_{n=1}^\infty
\frac{(-1)^n}{1-\frac{n^2}{a^2} } \cos(\frac{n\tau}{a})
\cos(\frac{n\s}{a}) \right),\\
\size
X_{\I \, 2}=\frac{\ell\sin(\pi a)}{\pi a} \left( \t +2a \sum_{n=1}^\infty
\frac{(-1)^n}{ n(1-\frac{n^2}{a^2}) } \sin(\frac{2n\tau}{a})
\cos(\frac{2n\s}{a}) \right),
\ea
\ee
where $-\pi a<\s<\pi a,$
and
\be
\ba{c}
\size
X_{\II \, 0}=2\ell\tau, \\
\size
X_{\II \, 1}=-\frac{\ell\sin(\pi a)}{\pi (1-a)} \left( 1+2\sum_{n=1}^\infty \frac{(-1)^n}{1-\frac{n^2}{(1-a)^2} }
\cos(\frac{n\tau}{(1-a)})  \cos(\frac{n\s}{(1-a)}) \right),\\
\size
X_{\II \, 2}=-\frac{\ell\sin(\pi a)}{\pi (1-a)} \left( \tau +2(1-a) \sum_{n=1}^\infty
\frac{(-1)^n}{n(1-\frac{n^2}{(1-a)^2}) } \sin(\frac{n\tau}{(1-a)})\cos(\frac{n\s}{(1-a)}) \right),
\ea
\ee
where $-\pi (1-a)<\s<\pi (1-a).$

Summing the series up, we obtain \be\la{31} X_{\I,\II \, \mu}(\s,\t
)=X_{\I,\II \, \m}^{+}(\s^{+})+X_{\I,\II \, \m}^{-}(\s^{-}) , \qquad
\s^{\pm}=\s\pm \t, \ee where \be\la{fl1} && X^\pm_{\I \, 0}=\pm
\frac{\ell}{2}a\s^\pm \ , \quad X^\pm_{\I \,
1}=\frac{\ell}{2}\bC_\I(\s^\pm ) \ , \quad X^\pm_{\I \, 2}=\pm
\frac{\ell}{2}\big[ \frac{\sin(a\pi )}{\pi}\s^\pm +\bS_\I(\s^\pm
)\big],
\\
\no
&&
\ba{cl}
\bC_\I(\z)=\cos(a\z ) \ , \quad \bS_\I (\z)=\sin(a\z)-\frac{\sin(a\pi)}{\pi}\z \quad
&{\rm for} \   0\leq \z < {\pi} \ , \\
\size \bC_\I(\z)=\cos(a\z-2a\pi) \ , \quad
\bS_\I(\z)=\sin(a\z-2a\pi)-\frac{\sin(a\pi )}{\pi}(\z -\pi) \quad
&{\rm for}	\ {\pi}\leq \z < 2\pi\ \ea \ee and \be\la{fl2} &&
X^\pm_{\II \, 0}=\pm \frac{\ell}{2}(1-a)\s^\pm \ , \quad X^\pm_{\II
\, 1}=\frac{\ell}{2}\bC_{\II}(\s^\pm )	 , \ \quad X^\pm_{\II \,
2}=\pm \frac{\ell}{2}\big[ -\frac{\sin(a\pi
)}{\pi}\s^\pm+\bS_{\II}(\s^\pm )\big],
\\
\no &&
\bC_{\II}(\z)=\cos((1-a)\z + a\pi) \ , \quad
\bS_{\II}(\z)=\sin((1-a)\z + a\pi)+\frac{\sin(a\pi)}{\pi}\z
\qquad {\rm for}  \ 0\leq \z < 2\pi .
\ee
In the expressions \rf{fl1} and \rf{fl2}, the
world-sheet parameters are rescaled as
\be\la{resc}
\ba{ll}
	\size {\rm fragment} \ \I \	 : \\
	\size {\rm fragment} \ \II:
\ea
\quad
\ba{c}
	\size  \t,\s \ \to \ a \t, a\s \\
	\size  \t,\s \ \to \ (1-a) \t, (1-a) \s.
\ea
\ee

The derivatives $X'_{\I,\II \, i}, \ i=1,2$ have discontinuities at
the points of splitting, i.e. at $\s^\pm =\pi$ for the fragment I
and $\s^\pm =0$ for the fragment II. These discontinuities show up
as an angular bending on the folded shape of the strings moving
along the strings as a function of $\tau$ (for more details see the
original paper \ci{IR}). Equations \rf{2} are satisfied at each
point on the string, in spite of the discontinuity. The
$\delta-$functions arising from the second derivative
$\del_{\s,\s}X_{\I,\II \, i}$ cancel with those coming from
$\del_{\t,\t} X_{\I,\II \, i},$ due to the chiral properties of
\rf{31}.

\section{Splitting of folded strings in AdS$_3.$}\la{sec4}

\renewcommand{\theequation}{4.\arabic{equation}}
 \setcounter{equation}{0}

In this section we discuss splitting of Gubser-Klebanov-Polyakov
folded spinning strings in AdS$_3$.

\subsection{Folded string in AdS$_3$}

The folded string solution in the AdS$_3$ in the embedding coordinates
read \ci{gkp}
\be\la{f1}
	Y_{05}=\cosh \r \ e^{i\k \t}, \qquad Y_{12}=\sinh \r \ e^{i\w \t},
\ee
where
\be\la{f2}
	\sinh \rho=\ell \ \sn\big[{\kappa \ell^{-1}	 \sigma},\ -\ell^2\big] \ ,
	\qquad
	\cosh \rho=\dn\big[{\kappa \ell^{-1}  \sigma},\ -\ell^2\big] \ ,
	\qquad
	\frac{w^2}{\kappa^2}= 1+ \frac{1}{\ell^2} \ .
\ee
Here $\sn[z,m]$ and $\dn[z,m]$ are the Jacobi elliptic functions,
$\ell$ defines the length of the string: $\sinh \r_{\max} = \ell.$

Expressions \rf{f2} are valid on the interval $0 \leq \s < \pih$ only.
To get the formal periodic solution
on the interval $ 0 \leq \s < 2 \pi$ one has to combine four stretches
of \rf{f2}:
\be\la{comb}
\ba{llllll}
	Y_{05}=\cosh \r(\s) e^{i\k \t} & & Y_{12}=\sinh\r(\s) e^{i\w \t}
	& & {\rm for} \quad \s \in [0,\pih)\\
	Y_{05}=\cosh \r(\pi - \s) e^{i\k \t} & & Y_{12}=\sinh \r(\pi - \s) e^{i\w \t}
	& & {\rm for} \quad \s \in [\frac{\pi}{2},\pi)\\
	Y_{05}=\cosh \r(\s - \pi) e^{i\k \t} & & Y_{12}=-\sinh \r(\s - \pi) e^{i\w \t}
	& & {\rm for} \quad \s \in [\pi,\frac{3\pi}{2})\\
	Y_{05}=\cosh \r(2\pi - \s) e^{i\k \t} & & Y_{12}=-\sinh \r(2\pi - \s) e^{i\w \t}
	& & {\rm for} \quad \s \in [\frac{3\pi}{2},2\pi).
\ea
\ee
and impose
\be
	Y_P(\s+2 \pi)=Y_P(\s).
\ee
The closed string periodicity conditions require
\be\la{f22}
	\kappa=\frac{2}{\pi} \ell \ \bK[-\ell^2].
\ee
The energy and spin are
\begin{equation}
\E= \frac{2}{\pi} \ell \ \bE[-\ell^2],
\qquad
\S= \frac{2}{\pi} \sqrt{1+\ell^2} \ (\bE[-\ell^2] - \bK[-\ell^2])
 \ .  \label{qdr}
\end{equation}
Here $\bK[z]$ and $\bE[z]$ are the complete elliptic integrals of
the first and second kinds, respectively.

The classical energy of the string in the limit of large spin is
\ci{gkp,bftt}\foot{ There is an elegant method to obtain expansion
for $\E(\S)$ in large or small $\S$ with arbitrary accuracy
\ci{paw}.} \be\la{lim0} E \simeq S+\frac{\sqrt{\lambda}}{\pi} \ln
\frac{S}{\sqrt{\lambda}}+  ... \,, \qquad
\frac{S}{\sqrt{\lambda}}\gg 1	\ . \ee

\subsection{Large-spin limit. Formal $\k=\w$ approximation.}\la{sec4.1}

There is a useful simplification of the solution \rf{f2},
when the spin of the folded string is large:
\be
\rho = \k \s\, ,
\qquad \k=\w \gg 1.
\label{tp}
\ee
This is a formal limit, as $\k \to \w$ implies $\ell \to \infty.$

\begin{figure}[ht]
\centerline{\includegraphics[scale=0.8]{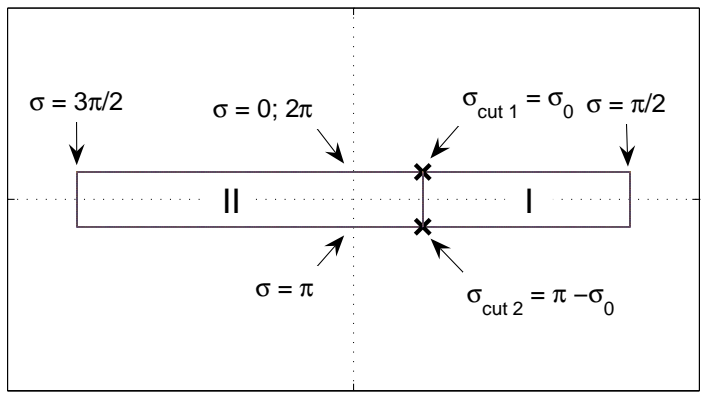}}
\caption{Splitting of the folded string in AdS$_3.$}
\label{fig1}\nonumber
\end{figure}

The energy and spin read
\be\la{ES0}
	\E = \S_{05} \simeq \frac{\k}{2 \pi} \pi +
		\frac{1}{4\pi} \ e^{\k \pi} \ ,
	\qquad
	\S = \S_{12} \simeq - \frac{\k}{2 \pi} \pi +
		\frac{1}{4\pi} \ e^{\k\pi} \ .
\ee Expansion of the classical energy in large $\S$ is consistent
with the one coming from \rf{qdr} in the first two orders\foot{ One
has to be careful using \rf{tp} for computing charges. It is easy to
see, that the absolute values of $\E$ and $\S$ in \rf{ES0}
approximately twice exceed those of \rf{qdr} taken at equal $\k.$
This inconsistency comes from the fact, that approximation \rf{tp}
is invalid close to the string ends \ci{bftt}, while the largest
contribution to the charges comes exactly from them.} \be\la{lim} E
\simeq S+\frac{\sqrt{\lambda}}{\pi} \ln \frac{S}{\sqrt{\lambda}}+
... \,, \qquad	\frac{S}{\sqrt{\lambda}}\gg 1	\ . \ee Any two
points on the string parameterized by $\s_1$ and $\s_2=\pi - \s_1$
coincide in the target space and their velocities agree at any given
time. Let us assume that at $\tau_0=0 $ the string splits into two
pieces. The cut occurs at $\r=\k \s_0,$ i.e. $\s_{cut \ 1}=\s_0$ and
$\s_{cut \ 2}=\pi - \s_0$ \be\la{frag1} \ba{ll}
	\size {\rm fragment} \ \I \	 : \quad \s \in (\s_0,\pi-\s_0)\\
	\size {\rm fragment} \ \II: \quad \s \in (0,\s_0) \cup (\pi - \s_0 , 2\pi),
\ea
\qquad 0 < \s_0 < \half.
\ee
Here without loss of generality $0<\s_0<\pih,$
i.e. the fragment I is always
smaller than the fragment II (see schematic plot in figure \ref{fig1}).

Approximation \rf{tp} is invalid
close to the string ends, thus we have to demand
\be\la{s0gg1}
	\pih-\s_0 \gg \frac{1}{\k}.
\ee

The charges $(\S_{\I,\II})_{PQ}$ of the cut fragments
read
\be\la{ch12}
\ba{c}
	\size (\S_{\I,\II})_{05} =
		\frac{\k}{2 \pi} \left( \frac{\pi}{2} \mp \s_0 \right)
		+ \frac{\sinh (\k \pi) \mp \sinh(2 \k \s_0)}{4 \pi}, \\
	\size
	 (\S_{\I,\II})_{12} =
		- \frac{\k}{2 \pi} \left( \frac{\pi}{2} \mp \s_0 \right)
		+ \frac{\sinh (\k \pi) \mp \sinh(2 \k \s_0)}{4 \pi},\\ [10pt]
	\size (\S_{\I,\II})_{02} =-(\S_{\I,\II})_{51}= \pm \frac{\cosh (\k \pi) - \cosh(2 \k \s_0)}{4 \pi},
	\qquad (\S_{\I,\II})_{01} =	 0, \qquad (\S_{\I})_{52} = 0.
\ea
\ee
They are conserved in the process of splitting
\be\la{col}
\ba{c}
	\size \E=\S_{05} = (\S_{\I})_{05}+(\S_{\II})_{05}, \qquad
	\S = (\S_{\I})_{12}+(\S_{\II})_{12},\\
	\size (\S_{in})_{02}=(\S_{\I})_{02}+(\S_{\II})_{02}, \qquad
	(\S)_{51}=(\S_{\I})_{51}+(\S_{\II})_{51}.
\ea
\ee

Spins of the fragments I and II given in \rf{ch12}
have non-Cartan components, as they are
written in the ``center-of-mass system'' of the
initial string (the coordinate system where
$\S_{PQ}$ of the string has Cartan components only).
It is more natural to analyze the fragments in their own
center-of-mass systems. Let us diagonalize $(\S_{\I,\II})_{PQ}.$

Performing the boost rotations independently for each string\foot{
Any rotation in the $Y_0Y_5,$ $Y_1Y_2,$ $Y_0Y_2$ or $Y_5Y_1$ would
result in $(\S_{\I,\II})_{01}$ and $(\S_{\I,\II})_{52}$ gaining nonzero
values. We are left only with boosts in $Y_0Y_1$ and $Y_5Y_2$
planes. } \be && \la{ab1} \left(\ba{c}\td Y_{\I,\II \, 0}\\ \td
Y_{\I,\II \, 1} \ea\right) = \left(
  \begin{array}{cc}
	\cosh \a_{\I,\II} & \sinh \a_{\I,\II} \\
	\sinh \a_{\I,\II} & \cosh \a_{\I,\II} \\
  \end{array}
\right)
\
\left( \ba{c} Y_{\I,\II \, 0} \\ Y_{\I,\II \, 1} \ea \right)
\\
&& \la{ab2}
\left(\ba{c} \td Y_{\I,\II \, 5} \\ \td Y_{\I,\II \, 2} \ea \right)
=
\left(
  \begin{array}{cc}
	\cosh \b_{\I,\II} & \sinh \b_{\I,\II} \\
	\sinh \b_{\I,\II} & \cosh \b_{\I,\II} \\
  \end{array}
\right) \ \left(\ba{c} Y_{\I,\II \, 5} \\ Y_{\I,\II \, 2} \ea
\right)
\ee
with the parameters ($\a_\I,\b_\I$ for the fragment I
and $\a_\II,\b_\II$ for the fragment II) \be\la{al}
	\size \a_{\I,\II} = \b_{\I,\II} = \mp \frac{\k}{2} \left(\frac{\pi}{2} \pm \s_0 \right),
\ee
we find the energies and spins of the cut fragments in
their own center-of-mass systems\foot{Making use of \rf{s0gg1}, we set
$\sinh(\k (\pih \mp \s_0)) \sim \half e^{\k (\pih \mp \s_0)}.$}
\be\la{III}
	\E_{\I,\II} \simeq
		\frac{\k}{2 \pi} \left(\frac{\pi}{2} \mp \s_0\right) + \frac{e^{\k (\pih \mp \s_0)}}{4\pi},
		\qquad
	\S_{\I,\II} \simeq
		-\frac{\k}{2 \pi} \left(\pih \mp \s_0\right) + \frac{e^{\k (\pih \mp \s_0)}}{4\pi}.
\ee
These expressions coincide with \rf{ES0} up to parameter definitions.
The expansions of the classical energies
$E_{\I,\II}(S_{\I,\II})$
in large spins obviously agree with \rf{lim}:
\be\la{ESIII}
E_{\I,\II} \simeq S_{\I,\II}+\frac{\sqrt{\lambda}}{\pi} \ln \frac{S_{\I,\II}}{\sqrt{\lambda}}+	 ... \,,
\qquad	\frac{S_{\I,\II}}{\sqrt{\lambda}}\gg 1.
\ee

Let us find the string solutions
describing the evolution of the cut fragments.

The evolution of the fragment I
is governed by the string equations \rf{3}
and \rf{2} with the initial conditions
at $\t_0=0$ (written in the center-of-mass
of the fragment):
\be\la{frI1}
\ba{c}
\size
	\td Y_{\I \, 0} = \cosh \left[ \k \left(\s -\frac{\pi}{4} - \frac{\s_0}{2} \right)\right],
	\qquad	\td Y_{\I \, 1} = \sinh \left[ \k \left(\s -\frac{\pi}{4} - \frac{\s_0}{2} \right)\right],\\
\size
	\td Y_{\I \, 5} = 0, \qquad \td Y_{I \, 2} = 0, \qquad
	\frac{\del}{\del \t} \td Y_{\I \, 0} = 0, \qquad \frac{\del}{\del \t} \td Y_{I \, 1} = 0,\\
\size
	\frac{\del}{\del \t} \td Y_{\I \, 5} = \k \cosh \left[ \k \left(\s -\frac{\pi}{4} - \frac{\s_0}{2} \right)\right],
	\qquad	\frac{\del}{\del \t} \td Y_{\I \, 2} =
		\k \sinh \left[ \k \left(\s -\frac{\pi}{4} - \frac{\s_0}{2} \right)\right]
\ea
\ee
for the interval $\s_0 < \s < \pih$ and the same expressions
with $\s \to \pi - \s$
for the interval $\pih < \s < \pi-\s_0.$

After rescaling of the world-sheet parameters $\s$ to
$\z$ in such a way that $\s_0 < \s < \pih  \to -\pih<\z<\pih:$
\be\la{sre}
	\s = \frac{\pi - 2\s_0}{2 \pi} \z + \frac{\pi}{4}+\frac{\s_0}{2}
	\quad {\rm and} \quad
	\t = \frac{\pi - 2\s_0}{2 \pi} \h,
\ee
we rewrite \rf{frI1} in the following form
\be\la{frI11}
\ba{c}
\size
	\td Y_{\I \, 0} = \cosh (\k_\I \z),
	\qquad	\td Y_{\I \, 1} = \sinh (\k_\I \z),
	\qquad \td Y_{\I \, 5} = 0,
	\qquad \td Y_{\I \, 2} = 0,
	\\
\size
	\frac{\del}{\del \h} \td Y_{\I \, 0} = 0, \qquad \frac{\del}{\del \h} \td Y_{I \, 1} = 0,
	\qquad
	\frac{\del}{\del \h} \td Y_{\I \, 5} = \k_\I \cosh (\k_\I \z),
	\qquad	\frac{\del}{\del \h} \td Y_{\I \, 2} = \k_\I \sinh (\k_\I \z),
	\\
\size
	\k_\I=\k \frac{\pi-2\s_0}{2\pi}.
\ea
\ee

Such boundary conditions are satisfied by
\be\la{roY}
	\size \td Y_{\I \, 05} = \cosh (\k_\I \z) e^{i \k_\I \h}, \qquad
	\size \td Y_{\I \, 12} = \sinh (\k_\I \z) e^{i \k_\I \h}.
\ee
That is the same as \rf{tp} up to parameter definitions.

For the fragment II we get similar result
\be\la{roY2}
	\size \td Y_{\II \, 05} = \cosh (\k_\II \z) e^{i \k_\II \h}, \qquad
	\size \td Y_{\II \, 12} = \sinh (\k_\II \z) e^{i \k_\II \h}, \qquad
	\k_\II=\k \frac{\pi+2\s_0}{2\pi}.
\ee

Making use of \rf{ES0} and \rf{III}
the following conservation laws
of the splitting may be derived\foot{
Here we used the relations
\be\no
	\E = \frac{\k}{2 \pi} \pi + \frac{1}{4\pi} e^{\k \pi}
	\quad \Ra \quad
	\ln \E = \k \pi - \ln 4\pi + 2 \pi \k e^{-\k \pi}
	\quad \Ra \quad
	\k\pi = \ln \E + \ln 4 \pi - \frac{2}{\E} \ln \E.
\ee
}
\be
	\E^{1-2/\E} = 4\pi \E_\I^{1-2/\E_\I} \ \E_\II^{1-2/\E_\II},
\qquad
	\S^{1+2/\S} = 4\pi \S_\I^{1+2/\S_\I} \ \S_\II^{1+2/\S_\II}.
\ee
The boost parameters \rf{al} may be expressed as
\be
	\a_{\I,\II}=\b_{\I,\II} \simeq
	\mp \ln \frac{\E^{1-2/\E}}{\E_{I,\II}^{1-2/\E_{\I,\II}}}
	\simeq \mp \ln \frac{\S^{1+2/\S}}{\S_{I,\II}^{1+2/\S_{\I,\II}}}.
\ee

Given \rf{tp}, \rf{roY}, \rf{roY2} and \rf{al} we see that when the
initial string described by the formal $\k=\w$ limit of the folded
string \rf{tp} splits into two pieces the world surface does not
change and the fragments are described by the solutions of the same
type as the parent string.

\vspace{20pt}

It is interesting to point out that \rf{tp} is not just a formal
approximation of the folded string profile (in the limit
$\k=\w \to \infty$), but a true solution of the string equations
\rf{3}, \rf{2} (with arbitrary values of $\k=\w$).
Strings of this type have a peculiar property. They may be
divided into an arbitrary number of fragments each of which is an
independent solution of the same type as \rf{tp}, simply boosted
from its center-of-mass. However, its stretches may not be
consistently glued to form a closed string. Such glued string would
have jumps of the first derivatives at the string ends $\r'(\s_{\rm
ends}+0) \neq \r'(\s_{\rm ends} -0),$ resulting in $\r''(\s_{\rm
ends})\sim \delta(\s \pm \s_{\rm ends}),$ and, consequently, would
not satisfy \rf{2}\foot{ In the flat space this inconsistency is
avoided due to the chiral properties of the solutions for the final
fragments (see above).}. The $\d-$functions arising on the r.h.s. of
the equations \rf{3} may be interpreted as point masses attached to
the string ends \ci{bn}.

\subsection{String with an arbitrary spin. The general case}\la{sec4.3}

In this section we discuss the most general case of splitting of
folded strings in AdS$_3.$ Starting with the folded string solution
in its exact form \rf{f1}, \rf{f2}, \rf{f22} and, following the
approach of section \ref{sec4.1}, we assume that the string splits
into two fragments (I and II) defined in \rf{frag1} and in figure
\ref{fig1}. Their charges $(\S_{\I,\II})_{PQ}$ read \be\la{c1}
\ba{c} \size
	(\S_{\I,\II})_{05} = \k \bC_{\I,\II} = \frac{\ell}{\pi} \left( \bE [-\ell^2] \mp
	\bE\left[ \am[\k \ell^{-1} \s_0,-\ell^2],-\ell^2\right] \right),\\[10pt]
\size
	(\S_{\I,\II})_{12} = \w \bS_{\I,\II} = - \frac{\sqrt{1+\ell^2}}{\pi^2} (\pi \mp 2\s_0)\bK[-\ell^2]
	+\frac{\sqrt{1+\ell^2}}{\pi^2} \left( \bE [-\ell^2] \mp
	\bE\left[ \am[\k \ell^{-1} \s_0,-\ell^2],-\ell^2\right] \right) ,\\
\size
	(\S_{\I,\II})_{02} = \w \bM_{\I,\II} = \pm \frac{1}{\pi} \ell \sqrt{1+\ell^2} \ \cn\left[\k \ell^{-1} \s_0,-\ell^2\right],
\qquad
	(\S_{\I,\II})_{15} = \k \bM_{\I,\II} = \pm \frac{1}{\pi} \ell^2 \ \cn\left[\k \ell^{-1} \s_0,-\ell^2\right] ,\\
\size
	(\S_{\I,\II})_{01} = (\S_{\I,\II})_{52} =0 \ ,
\ea
\ee
where $\bE[z,m]$ is the incomplete elliptic integral
of the second kind.
We want to find a coordinate systems where the non-Cartan
components of the spins vanish
and find the energies and spins of the
final fragments. $(\S_{\I,\II})_{PQ}$ may be
diagonalized by boosts
in the $Y_0Y_1$ and $Y_5Y_2$ planes \rf{ab1}, \rf{ab2}
with parameters
\foot{
Vanishing of the non-Cartan components of the spins
implies
\be\la{ESg2}
&& \no {\mathbb{M}}_{\I,\II} (\k + \w) \cosh(\a_{\I,\II} + \b_{\I,\II})
+ (\k {\mathbb{C}}_{\I,\II} + \w {\mathbb{S}}_{\I,\II}) \sinh(\a_{\I,\II} + \b_{\I,\II}) = 0\\
&& \no {\mathbb{M}}_{\I,\II} (\k - \w) \cosh(\a_{\I,\II} - \b_{\I,\II})
+ (\k {\mathbb{C}}_{\I,\II} - \w {\mathbb{S}}_{\I,\II}) \sinh(\a_{\I,\II} - \b_{\I,\II}) = 0.
\ee
That leads to \rf{ab4}.
}
\be\la{ab4}
\ba{c}
\size
	\sinh(\a_{\I,\II} + \b_{\I,\II}) = - \frac{\bM_{\I,\II} (\w + \k)}
	{\sqrt{ (\k \bC_{\I,\II} + \w \bS_{\I,\II})^2 - \bM_{\I,\II}^2 (\w + \k)^2}} \\
\size
	\sinh(\a_{\I,\II} - \b_{\I,\II}) = \frac{\bM_{\I,\II} (\w - \k)}
	{\sqrt{ (\k \bC_{\I,\II} - \w \bS_{\I,\II})^2 - \bM_{\I,\II}^2 (\w - \k)^2}},
\ea
\ee
where
\be\la{csm3}
\ba{c}
\size
	\bC_{\I,\II} =
	\frac{\ell}{\k \pi} \left( \bE [-\ell^2] \mp
	\bE\left[ \am[\k \ell^{-1} \s_0,-\ell^2],-\ell^2\right] \right),
\qquad
	\bS_{\I,\II} = - \frac{\pi \mp 2\s_0}{2} + \bC_{\I,\II}\\[8pt]
\size
	\bM_{\I,\II} = \pm \frac{\ell^2}{\k \pi} \ \cn\left[\k \ell^{-1} \s_0,-\ell^2\right],
	\qquad \kappa=\frac{2}{\pi} \ell \ \bK[-\ell^2].
\ea
\ee
Then the energies and spins of the cut fragments read
\be\la{ESiii}
\ba{c}
\size
	\E_{\I,\II} =\half \sqrt{ (\k \bC_{\I,\II} + \w \bS_{\I,\II})^2 - \bM_{\I,\II}^2 (\w + \k)^2}
		+ \half \sqrt{ (\k \bC_{\I,\II} - \w \bS_{\I,\II})^2 - \bM_{\I,\II}^2 (\w - \k)^2}
	\\
\size
	\S_{\I,\II} = \half \sqrt{ (\k \bC_{\I,\II} + \w \bS_{\I,\II})^2 - \bM_{\I,\II}^2 (\w + \k)^2}
		- \half \sqrt{ (\k \bC_{\I,\II} - \w \bS_{\I,\II})^2 - \bM_{\I,\II}^2 (\w - \k)^2}.
\ea
\ee
These relations parametrically encode the conservation laws of
splitting, e.g. $E(E_\I,E_\II),$ $S(S_\I,S_\II),$ etc.

The evolution of the fragments I and II is governed by the string
equations \rf{3} and \rf{2} with the boundary conditions given by
the initial string \rf{f1}, \rf{f2}, \rf{f22} on the intervals
\rf{frag1} at $\t_0=0$. Due to the complexity of the folded string
profile \rf{f2}, we are unable to find solutions to these equations. However, we
could describe the evolution qualitatively based on the result of
section \ref{sec4.1} and section \ref{sec3}, in the limit of large ---
but not infinitely large (as in \rf{tp}) --- spin, so long as the cut occurs
far enough from the string ends for $\s_0$ to satisfy $\k(\pi/2 -
\s_0)\gg 1.$ In this case, one should expect the final pieces have
almost the standard folded shape \rf{f1}, \rf{f2}, \rf{f22}, which
is disturbed by a kink moving along the string, similar to that
observed in flat Minkowski space, see section \ref{sec3} and
\ci{IR}.
The kink is a ``correction'' to the ``leading'' folded shape of the
cut fragments, thus the angle of bending has to depend on the position
of the kink. It may be substantial at the string ends but must
be small close to the center.

\section{Splitting of strings in AdS$_3 \times$S$^5$.}\la{sec5}

\renewcommand{\theequation}{5.\arabic{equation}}
 \setcounter{equation}{0}

In this section we generalize the results
for the splitting of the folded string in AdS$_3$ to
AdS$_3 \times$S$^5,$ including into consideration
circular rotations and windings in S$^5.$

Let us consider the string solution
having the folded shape in the AdS$_3$ and
the circular one with windings in S$^3:$
\be
\la{fg1}
\ba{c}
\size
	 Y_{05}=\cosh \r \ e^{i\k \t}, \qquad Y_{12}=\sinh \r \ e^{i\w \t},
	\\
\size
	X_{12}=a e^{i(\n \t + m \s)}, \qquad X_{34}=b e^{i(\n \t - m \s)},
	\qquad a^2 + b^2 =1, \qquad m \in N,
\ea
\ee
where
\be\la{f3}
\ba{c}
\size
	\sinh \rho=\td \ell \ \sn\big[{\td \kappa \td \ell^{-1}	 \sigma},\ -\td \ell^2\big],
	\qquad
	\td \k=\frac{2}{\pi} \td \ell \ \bK[-\td \ell^2] , \\
\size
	\frac{\td \w^2}{\td \k^2} = 1 + \frac{1}{\tell^2},
	\qquad
	\td \w^2 = \w^2 - (\n^2 + m^2) ,
	\qquad
	\td \k^2 = \k^2 - (\n^2 + m^2).
\ea \ee
Comparing that with \rf{f1}, \rf{f2} and \rf{f22}, we see
that the only result of accounting for the S$^3$ part is redefinition
of $\k,\w \to \td k,\td w.$ That is also true if one adds other
spins and windings in S$^5.$

Combining together four stretches of \rf{fg1}, each of which is
valid on the interval $0 \leq \s < \pih,$ we obtain a periodic
solution on the interval $ 0 \leq \s < 2 \pi.$ Its classical energy
and spins read \begin{equation}\la{ESjj} \ba{c} \size \E_\J=
\frac{2}{\pi} \ \frac{\k}{\td \k} \ \td \ell \ \bE[-\td \ell^2] =
\sqrt{ \frac{4}{\pi^2} \td \ell^2 + \frac{m^2 + (\J_1 + \J_2))^2 }{
\bK^2[-\td \ell^2] }} \ \bE[-\td \ell^2] \ ,
\\
\size
\S_\J= \frac{2}{\pi} \ \frac{\w}{\td \w} \
\sqrt{1+\td \ell^2} \ (\bE[-\td \ell^2] - \bK[-\td \ell^2])
=\sqrt{ \frac{4}{\pi^2} (1+\td \ell^2) + \frac{ m^2 + (\J_1 + \J_2))^2 }{ \bK^2[-\td \ell^2] }}
\ (\bE[-\td \ell^2] - \bK[-\td \ell^2]),
\\
\size
\J_1 = a^2 \n, \qquad \J_2 = b^2 \n, \qquad \n = \J_1 + \J_2 \ .
\ea
\end{equation}

Following the approach of section \ref{sec4}, first, we consider the
limit of the string with large spin in AdS$_3$. Then the AdS-part of
the solution \rf{f3} may be approximated by \be \rho=\sqrt{\k^2 -
(\n^2+m^2)} \ \sigma = \td \k \s\, , \qquad \k=\w \, , \qquad \tk
\gg 1. \label{tps} \ee This is a formal limit as $\k \to \w$ implies
$\tell \to \infty.$

The energy and AdS-spin of the string
read
\be\la{ESs}
	\E_\J = \S_{05} \simeq \frac{\k}{2 \pi} \pi +
		\frac{\k}{4\pi \tk} \ e^{\td \k \pi},
	\qquad
	\S_\J = \S_{12} \simeq - \frac{\k}{2 \pi} \pi +
		\frac{\k}{4\pi \tk} \ e^{\tk \pi}.
\ee
Spins in S$^3$ are unaffected by the limit.

Two points on the string parameterized by $\s_1$ and $\s_2$
coincide in the target space and their
velocities agree, if $\s_1=\s_0,$ $\s_2=\pi - \s_0$ and
\be\la{s0}
	\s_0 = \left(\half - \frac{n}{m}\right) \pi, \qquad n \in N,
	\qquad {\rm if } \ m\neq 0
\ee or for arbitrary $\s_0$ if $m=0.$ The string is not folded in
AdS$_3\times$S$^3,$ when $m \neq 0.$

Approximation \rf{tps} is invalid
close to the string ends, thus we have to demand
\be\la{s0lim}
	\pih-\s_0 \gg \frac{1}{\tk}
\ee
for the coordinates of the
cut ($\s_{{\rm cut} \, 1}=\s_0$ and
$\s_{{\rm cut} \, 2}=\pi - \s_0$).

The charges $(\S^\J_{\I,\II})_{PQ}$ of the cut fragments read
\be\la{ch1s}
\ba{c}
	\size (\S^\J_{\I,\II})_{05} =
		\frac{\k}{2 \pi} \left( \frac{\pi}{2} \mp \s_0 \right)
		+ \frac{\k}{\tk} \ \frac{\sinh(\tk \pi) \mp \sinh(2 \tk \s_0)}{4 \pi}, \\
	\size
	 (\S^\J_{\I,\II})_{12} =
		- \frac{\k}{2 \pi} \left( \frac{\pi}{2} \mp \s_0 \right)
		+ \frac{\k}{\tk} \ \frac{\sinh(\tk \pi) \mp \sinh(2 \tk \s_0)}{4 \pi},\\ [10pt]
	\size (\S^\J_{\I,\II})_{02} = - (\S^\J_{\I,\II})_{51}
	= \pm \frac{\k}{\tk} \ \frac{\cosh(\tk \pi) - \cosh(2 \tk \s_0)}{4 \pi},
	\qquad (\S^\J_{\I,\II})_{01} =	 0, \qquad (\S^\J_{\I})_{52} = 0, \\[10pt]
	\size
	(\J_{\I,\II})_1=a^2 \, \frac{\n}{2\pi}(\pi \mp 2\s_0),
	\qquad (\J_{\I,\II})_2=b^2 \, \frac{\n}{2\pi}(\pi \mp 2\s_0).
\ea
\ee
They are conserved in the process of splitting
\be\la{cols}
\ba{c}
	\size \E^\J=(\S^\J_{\I})_{05}+(\S^\J_{\II})_{05}, \qquad
	\S^\J = (\S^\J_{\I})_{12}+(\S^\J_{\II})_{12},\\
	\size \S^\J_{02}=(\S^\J_{\I})_{02}+(\S^\J_{\II})_{02}, \qquad
	\S^\J_{51}=(\S^\J_{\I})_{51}+(\S^\J_{\II})_{51}, \\
	\size \J_1 = (\J_\I)_1+(\J_\II)_1, \qquad
	\J_2 = (\J_\I)_2+(\J_\II)_2.
\ea
\ee
It is natural to transform \rf{ch1s} to the center-of-mass
systems of the final strings and explicitly find their energies and spins.
$(\S^\J_{\I,\II})_{PQ}$ may be
diagonalized by boosts
in the $Y_0Y_1$ and $Y_5Y_2$ planes \rf{ab1}, \rf{ab2}
with parameters 
\be\la{abj}
	\size \a^\J_{\I,\II} = \b^\J_{\I,\II} = \mp \frac{\tk}{2} \left(\frac{\pi}{2} \pm \s_0 \right).
\ee
We obtain the energies and AdS-spins of the fragments
in the form\foot{Making use of \rf{s0lim}, we set
$\sinh (\tk (\pih \mp \s_0)) \to \half e^{\tk (\pih \mp \s_0)}$.}
\be\la{ESj}
	\E^\J_{\I,\II} =
		\frac{\k}{2 \pi} \left(\frac{\pi}{2} \mp \s_0\right) + \frac{\k}{\tk} \ \frac{e^{\tk (\pih \mp \s_0)}}{4\pi},
		\qquad
	\S^\J_{\I,\II} =
		-\frac{\k}{2 \pi} \left(\pih \mp \s_0\right) + \frac{\k}{\tk} \ \frac{e^{\tk (\pih \mp \s_0)}}{4\pi}.
\ee

The evolution of the fragments (in the own center-of-mass system for each fragment) is described by
\be\la{roYs}
\ba{c}
	\size (\td Y_{\I,\II})_{05} = \cosh (\tk_{\I,\II} \z) e^{i \k_{\I,\II} \h} \qquad
	(\td Y_{\I,\II})_{12} = \sinh (\tk_{\I,\II} \z) e^{i \k_{\I,\II} \h} \\
	\size (\td X_{\I,\II})_{12}=a e^{i(\n_{\I,\II} \h + m_{\I,\II} \z)} \qquad
	 (\td X_{\I,\II})_{34}=b e^{i(\n_{\I,\II} \h - m_{\I,\II} \z)},
\ea
\ee
where
\be
	\k_{\I,\II}=\k \frac{\pi_0 \mp 2\s_0}{2\pi}, \qquad
	\tk_{\I,\II}=\tk \frac{\pi_0 \mp 2\s_0}{2\pi}, \qquad
	\n_{\I,\II} = \n \frac{\pi \mp 2\s_0}{2 \pi}, \qquad
	m_{\I,\II} = m \frac{\pi \mp 2\s_0}{2 \pi} 
\ee
and $\s_0$ satisfy \rf{s0} if $m \neq 0.$
Note, that while the AdS part of \rf{roYs}
is just a large-spin approximation, the solution for the
S$^3$ part is exact.

Given \rf{tps}, \rf{roYs} and \rf{abj} we see that, when the initial
string, described by the formal $\k=\w$ limit of the string \rf{tps}
in AdS$_3\times$S$^3,$ splits into two pieces the world surface does
not change and the fragments are described by the solutions of the
same type as the parent string.

In the general case, starting from the exact solution
in AdS$_3\times$S$^3$ in the form \rf{fg1}, \rf{f3},
we obtain the following expressions for
the charges of the cut fragments
(in the center-of-mass of the initial string):
\be\la{c5}
\ba{c}
\size
	(\S^\J_{\I,\II})_{05} = \half \E_\J \mp
	\sqrt{ \frac{1}{\pi^2} \td \ell^2 + \frac{\J^2 }{ 4 \bK^2[-\td \ell^2] }}
	\bE\left[ \am[\tk \td \ell^{-1} \s_0,-\td\ell^2],-\td\ell^2\right],\\[10pt]
\size
	(\S^\J_{\I,\II})_{12} = \half \S_\J
	\mp \sqrt{ \frac{1}{\pi^2} (1+\td \ell^2) + \frac{\J^2 }{ 4 \bK^2[-\td \ell^2] }}
	\left( \bE\left[ \am[\tk \td\ell^{-1} \s_0,-\td\ell^2],-\td\ell^2\right]
	- \frac{2}{\pi}\s_0 \bK[-\td\ell^2] \right) ,\\
\size
	(\S^\J_{\I,\II})_{02} = \pm \w \frac{\tell^2 \ \cn\left[\tk \tell^{-1} \s_0,-\tell^2\right]}{\k \pi },
\qquad
	(\S^\J_{\I,\II})_{15} = \pm \frac{\tell^2 \ \cn\left[\tk \tell^{-1} \s_0,-\tell^2\right]}{\pi} ,\\
\size
	(\S^\J_{\I,\II})_{01} = (\S^\J_{\I,\II})_{52} =0,\\
\size
	(\J_{\I,\II})_1=a^2 \, \frac{\n}{2\pi}(\pi \mp 2\s_0),
	\qquad (\J_{\I,\II})_2=b^2 \, \frac{\n}{2\pi}(\pi \mp 2\s_0),
\ea \ee where $\E_\J$ and $\S_\J$ are defined in \rf{ESjj}. That may
be transformed to the center-of-mass systems of the final states by
the boosts in the $Y_0Y_1$ and $Y_5Y_2$ planes \rf{ab1}, \rf{ab2}
with parameters \be \ba{c} \size
	\sinh(\a^\J_{\I,\II} + \b^\J_{\I,\II}) = - \frac{\td\bM_{\I,\II} (\w + \k)}
	{\sqrt{ (\k \td\bC_{\I,\II} + \w \td\bS_{\I,\II})^2 - \td\bM_{\I,\II}^2 (\w + \k)^2}} \\
\size
	\sinh(\a^\J_{\I,\II} - \b^\J_{\I,\II}) = \frac{\td\bM_{\I,\II} (\w - \k)}
	{\sqrt{ (\k \td\bC_{\I,\II} - \w \td\bS_{\I,\II})^2 - \td\bM_{\I,\II}^2 (\w - \k)^2}},
\ea
\ee
where $\td\bC_{\I,\II}, \ \td\bS_{\I,\II}$ and $\td\bM_{\I,\II}$
are given by \rf{csm3} with $\ell$ replaced for $ \tell.$

The general expressions for the energies and
spins of the fragments read
\be\la{esjj}
\no
	& \E^\J_{\I,\II} = \half \sqrt{ (\k \td\bC_{\I,\II} + \w \td\bS_{\I,\II})^2 - \td\bM_{\I,\II}^2 (\w + \k)^2}
		+ \half \sqrt{ (\k \td\bC_{\I,\II} - \w \td\bS_{\I,\II})^2 - \td\bM_{\I,\II}^2 (\w - \k)^2} &
	\\
\no
	& \S^\J_{\I,\II} = \half \sqrt{ (\k \td\bC_{\I,\II} + \w \td\bS_{\I,\II})^2 - \td\bM_{\I,\II}^2 (\w + \k)^2}
		- \half \sqrt{ (\k \td\bC_{\I,\II} - \w \td\bS_{\I,\II})^2 - \td\bM_{\I,\II}^2 (\w - \k)^2} &
	\\
\no
	& (\J_{\I,\II})_1 = a^2 \, \frac{\n}{2\pi}(\pi \mp 2\s_0),
	\qquad (\J_{\I,\II})_2=b^2 \, \frac{\n}{2\pi}(\pi \mp 2\s_0). &
\ee
These relations parametrically encode the conservation laws of
splitting.

The evolution of the fragments I and II is governed by the
string equations \rf{3} and \rf{2} with the boundary conditions
given by the initial string \rf{fg1}, \rf{f3} on the intervals
\rf{frag1} at $\t_0=0$ with $\s_0$ satisfying \rf{s0}. The solutions
describing the profiles of the fragments consist of AdS- and
S$^3$-parts. The expressions for the S$^3$-parts presented in
\rf{roYs}, but we are unable to find the exact expressions for the
AdS-parts, due to the complexity of \rf{f3}. Up to parameter
definitions, the AdS-parts coincide with the solutions describing
the splitted fragments of the folded string in pure AdS$_3.$
This is based on the fact that the only result of accounting for
the S$^5$ is redefinition of $\k,\w \to \td k,\td w$ and
discretizing of $\s_0,$ if any.

\section*{Concluding remarks}

In this paper we have investigated splitting of folded spinning
strings in AdS$_3$ and its generalization to include circular
rotations and windings in S$^5.$ We computed the energies and spins
of products of splitting and showed that in the case of splitting
of strings with large AdS-spins (which is of greatest interest in
the context of AdS/CFT duality) the cut fragments are described by
the solutions very similar to the initial string. The complexity of the
exact folded string profile prevents us from finding the evolution
of the final fragments by solving the string equations with
boundary conditions given by the initial string. However, one hopes
that this might be reachable ``indirectly'' by applying the finite
gap technique (see \ci{sakura,vicedo} for reviews). The profiles of
the cut fragments are known at the moment of splitting, thus we can
find the full set of the conserved charges for them, including the
higher ones. This uniquely determines the algebraic surface
which, being explicitly constructed, would allow the determination of
the string profiles. Implementation of such an approach is
promising, but quite complicated. It requires detailed
investigation.

\section*{Acknowledgments}

I am grateful to Arkady Tseytlin for the most valuable discussions
and suggestion to look at the strings splitting problem, to
Jorge Russo for sharing his Mathematica files and to David Weir
for the help with proofreading.
This work is supported by a grant of the
Dynasty Foundation and in part by the grant Scientific Schools SS---4142.2010.2.

\end{document}